\documentclass[12pt,a4paper]{article}
\pdfoutput=1

\usepackage[top=3.3cm, bottom=2.5cm, left=2.9cm, right=2.9cm]{geometry}

\usepackage[pdftex,breaklinks,colorlinks,
linkcolor=blue,
citecolor=blue,
urlcolor=blue]{hyperref}

\usepackage[T1]{fontenc}

\usepackage[font=small,labelfont=bf,width=0.85\textwidth]{caption}
\usepackage{amsmath}
\usepackage{amsfonts}
\usepackage{amsthm}
\usepackage{amssymb}

\usepackage{graphicx}
\usepackage{xcolor}

\usepackage{mathptmx}

\usepackage[round]{natbib}

\newtheorem{thm}{\indent Theorem}

\newtheorem{claim}{Claim}

\newenvironment{myproof}%
{{\indent\mbox{\em Proof.\ }}}%
{\hfill$\blacksquare$\vspace{1pt}}

\newenvironment{claimproof}%
{{\noindent\mbox{\em Proof.\ }}}%
{\hfill\qed\vspace{0pt}}

\title{On circular-arc graphs having a model\\with no three arcs covering the circle\footnote{This paper originally appeared in proceedings of the XVI Congreso Latino-Iberoamericano de Investigación Operativa and the XLIV Simp\'{o}sio Brasileiro de Pesquisa Operacional, September 24-28, 2012, Rio de Janeiro, Brazil. \emph{Anais do XLIV Simp\'{o}sio Brasileiro de Pesquisa Operacional}, SOBRAPO, Rio de Janeiro, Brazil, 2012, pages 4093--4104. URL: \url{http://www.din.uem.br/sbpo/sbpo2012/pdf/arq0518.pdf}} \footnote{E-mail addresses: L.N. Grippo (\href{mailto:lgrippo@ungs.edu.ar}{lgrippo@ungs.edu.ar}) and M.D. Safe (\href{mailto:msafe@ungs.edu.ar}{msafe@ungs.edu.ar})}}

\author{Luciano N. Grippo \and Mart\'in D. Safe}
\date{\begin{small} Instituto de Ciencias, Universidad Nacional de General Sarmiento\\Los Polvorines, Buenos Aires, Argentina \end{small}}

\begin{document}

\maketitle
\abstract{An \emph{interval graph} is the intersection graph of a finite set of intervals on a line and a \emph{circular-arc graph} is the intersection graph of a finite set of arcs on a circle. While a forbidden induced subgraph characterization of interval graphs was found fifty years ago, finding an analogous characterization for circular-arc graphs is a long-standing open problem. In this work, we study the intersection graphs of finite sets of arcs on a circle no three of which cover the circle, known as \emph{normal Helly circular-arc graphs}. Those circular-arc graphs which are minimal forbidden induced subgraphs for the class of normal Helly circular-arc graphs were identified by Lin, Soulignac, and Szwarcfiter, who also posed the problem of determining the remaining minimal forbidden induced subgraphs. In this work, we solve their problem, obtaining the complete list of minimal forbidden induced subgraphs for the class of normal Helly circular-arc graphs.}


\section{Introduction}

The intersection graph of a finite family of sets has one vertex representing each member of the family, two vertices being adjacent if and only if the members of the family they represent have nonempty intersection. An \emph{interval graph} is the intersection graph of a finite set of intervals on a line. Fifty years ago, \citet{L-B-interval-AT} found their celebrated characterization of interval graphs by minimal forbidden induced subgraphs. An interesting special case of interval graphs are the \emph{unit interval graphs}, which are the intersection graphs of finite sets of all closed (or all open) intervals having the same length, and for which there is a forbidden induced subgraph characterization due to \citet{Rob-indiff} \citep[see also][]{F-M-openclosed}.

A \emph{circular-arc graph} is the intersection graph of a finite set of arcs on a circle. Despite their similarity in definition to interval graphs, characterizing circular-arc graphs by forbidden induced subgraphs is a long-standing open problem (\citealp[see][p.\ 54]{HD-combgeom}; \citealp{Klee-circ-arc}). \citet{Tuc-matrix} pioneered the study of circular-arc graphs and some important subclasses, like \emph{unit circular-arc graphs} (defined analogously to unit interval graphs) and \emph{proper circular-arc graphs}, which are those intersection graphs of finite sets of arcs on a circle such that none of the arcs is contained in another of the arcs. Moreover, \citet{Tuc-struc} found the minimal forbidden induced subgraph characterizations of both unit circular-arc graphs and proper circular-arc graphs. Since then, the problem of characterizing circular-arc graphs and some of its subclasses by forbidden induced subgraphs or some other kinds of obstructions has attracted considerable attention \citep{BangJensenHell94,B-D-G-S-circ-arc,F-H-H-listhom,H-H-intbigraphs,J-L-M-S-S-HCA,L-S-S-properHelly,L-S-S-normal,TrotterMoore76}.

We say that a set of arcs on a circle \emph{covers the circle} if the arcs of the set collectively cover every point of the circle. It is easy to see that every circular-arc graph is the intersection graph of a finite set of arcs no single arc of which covers the circle. Intersection graphs of finite sets of arcs on a circle no two arcs of which cover the circle are known as \emph{normal circular-arc graphs}. The class of normal circular-arc graphs properly contains the class of proper circular-arc graphs as shown by \citet{Tuc-struc} and was studied in the context of co-bipartite graphs by \citet{H-H-intbigraphs}. Some partial characterizations by minimal forbidden induced subgraphs for the class of normal circular-arc graphs are known \citep[see][especially Section~5]{B-D-G-S-circ-arc}.

In this work, we study the intersection graphs of finite sets of arcs on a circle no three arcs of which cover the circle, known as \emph{normal Helly circular-arc graphs} \citep{L-S-S-properHelly}. Notice that, for any set of arcs on a circle having at least three arcs, the property of not having three arcs covering the circle precludes also the existence of fewer than three arcs covering the circle. If $\mathcal A$ is a set of arcs on a circle, then: (i) $\mathcal A$ is said \emph{normal} if it has no two arcs covering the circle, (ii) $\mathcal A$ is said \emph{Helly} if every nonempty subset of $\mathcal A$ consisting of pairwise intersecting arcs has nonempty total intersection, and (iii) $\mathcal A$ is said \emph{normal Helly} if $\mathcal A$ is both normal and Helly. In turns out that normal Helly circular-arc graphs can be defined as the intersection graphs of finite normal Helly sets of arcs on a circle. Indeed, it follows from Theorem~1 of \citet{L-S-circ-arc} that this definition of normal Helly circular-arc graphs is equivalent to the one we use along this work (i.e., the intersection graphs of finite sets of arcs on a circle no three arcs of which cover the circle). 

Some previous works related to normal Helly circular-arc graphs are the following. \citet{Tuc-col} gave an algorithm that outputs a proper coloring of any given normal Helly circular-arc graph using at most $3\omega/2$ colors, where $\omega$ denotes the maximum size of a set of pairwise adjacent vertices. In~\citet{L-S-S-cliqueca}, normal Helly circular-arc graphs arose naturally when studying convergence of circular-arc graphs under the clique operator. The \emph{boxicity} of a graph $G$ is the minimum $k$ such that $G$ is the intersection graph of a family of $k$-dimensional boxes (i.e., of Cartesian products of $k$ closed intervals); it was shown by \cite{B-SC-boxicity} that normal Helly circular-arc graphs have boxicity at most $3$.

Recently, \citet*{L-S-S-normal} undertook a thorough study of normal Helly circular-arc graphs, drawing many parallels between these graphs and interval graphs. In that work, they determined all those circular-arc graphs which are minimal forbidden induced subgraphs for the class of normal Helly circular-arc graphs and posed the problem of finding the remaining minimal forbidden subgraphs for the class of normal Helly circular-arc graphs (i.e., those which are not circular-arc graphs). In this work, we solve their problem, providing the complete list of minimal forbidden induced subgraphs for the class of normal Helly circular-arc graphs.

\section{Preliminaries}

All graphs in this work are finite, undirected, and have no loops and no parallel edges. The vertex and edge sets of a graph $G$ will be denoted by $V(G)$ and $E(G)$, respectively. We denote by $\overline G$ the \emph{complement} of $G$, by $N_G(v)$ the \emph{neighborhood} of a vertex $v$ in $G$, and by $N_G[v]$ its \emph{closed neighborhood} $N_G(v)\cup\{v\}$. Vertex $v$ is \emph{isolated} if $N_G(v)=\emptyset$ and \emph{universal} if $N_G[v]=V(G)$. The subgraph \emph{induced} by a set of vertices $S$ is denoted by $G[S]$. If $H$ is an induced subgraph of $G$, we say that $G$ \emph{contains an induced} $H$. Paths in this work are meant to have at least one vertex. An $a,b$-\emph{path} is a path whose \emph{endpoints} are $a$ and $b$; the remaining vertices of the path are the \emph{interior vertices}. A \emph{chord} of a path or cycle $Z$ is any edge not in $Z$ joining two vertices of $Z$. A \emph{chordless path} is a path having no chords and a \emph{chordless cycle} is a cycle on four or more vertices having no chords. We denote by $P_n$ (resp.\ $C_n$) the chordless path (resp.\ cycle) on $n$ vertices. A graph is \emph{chordal} if it has no chordless cycle. A \emph{clique} is a set of pairwise adjacent vertices. Two vertices are in the same \emph{component} of a graph $G$ if there is a path joining them in $G$. If $S$ is a set, we denote its cardinality by $\vert S\vert$. For standard notation and terminology not defined here, we refer to \cite{West-2ed}.

Let $\mathcal G$ be a graph class. A graph $H$ is a \emph{forbidden induced subgraph} of $\mathcal G$ if no member of $\mathcal G$ contains an induced $H$. A class $\mathcal G$ of graphs is \emph{hereditary} if every induced subgraph of every member of $\mathcal G$ is also a member of $\mathcal G$. If $H$ is a forbidden induced subgraph of a hereditary graph class $\mathcal G$, then $H$ is a \emph{minimal forbidden induced subgraph} of $\mathcal G$ if every induced subgraph of $H$ different from $H$ is a member of $\mathcal G$. Clearly, a hereditary graph class is completely determined by the its minimal forbidden induced subgraphs: the graph class consists exactly of those graphs containing no induced minimal forbidden induced subgraph. The following celebrated result gives the complete list of minimal forbidden induced subgraphs for the class of interval graphs.

\begin{figure}[ht!]
\centering
  \includegraphics{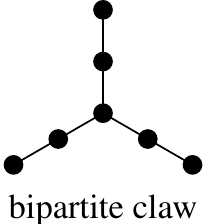}\quad\includegraphics{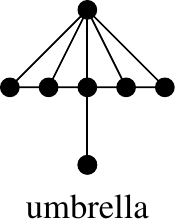}\quad\includegraphics{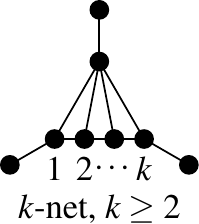}
    \quad\includegraphics{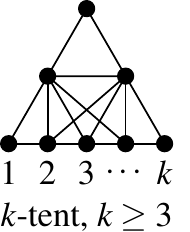}\quad\includegraphics{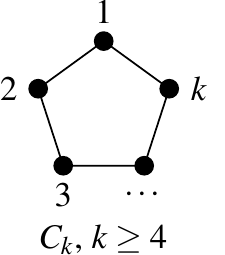}
\caption{All minimal forbidden induced subgraphs for the class of interval graphs}\label{fig:LekkerkerkerBoland}
\end{figure}

\begin{figure}
\centering
\includegraphics{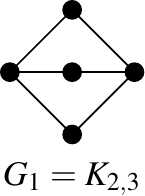}\quad\includegraphics{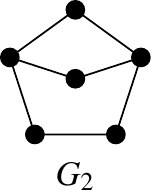}\quad\includegraphics{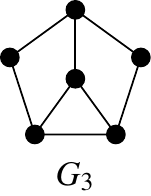}\quad\includegraphics{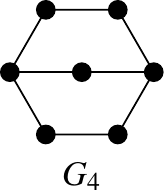}\\
\medskip\includegraphics{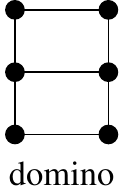}\quad\includegraphics{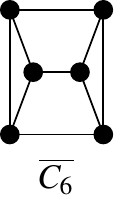}\quad\includegraphics{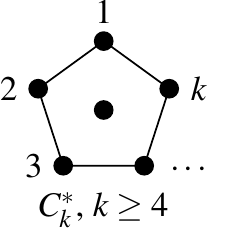}\quad\includegraphics{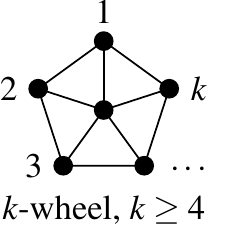}
\caption{Some minimal forbidden induced subgraphs for the class of normal Helly circular-arc graphs}\label{fig:technical}
\end{figure}

\begin{thm}[\citealp{L-B-interval-AT}]\label{thm:Lekkerkerker-Boland} The minimal forbidden induced subgraphs for the class of interval graphs are: bipartite claw, umbrella, $k$-net for every $k\geq 2$, $k$-tent for every $k\geq 3$, and $C_k$ for every $k\geq 4$ (see Figure~\ref{fig:LekkerkerkerBoland}).\end{thm}
In contrast with the situation for interval graphs, the problem of characterizing circular-arc graphs by forbidden induced subgraphs is still open. Some minimal forbidden induced subgraphs for the class of circular-arc graphs are $G_1$, $G_2$, $G_3$, $G_4$, domino, $\overline{G_6}$, and $C_k^*$ for every $k\geq 4$, where $C_k^*$ denotes the graph that arises from $C_k$ by adding an isolated vertex (see~Figure~\ref{fig:technical}); it follows, for instance, from our main result (Theorem~\ref{thm:main}) that these graphs are also minimal forbidden induced subgraphs for the class of normal Helly circular-arc graphs. For each $k\geq 4$, the graph \emph{$k$-wheel}, that arises from $C_k$ by adding a universal vertex (see~Figure~\ref{fig:technical}), is a circular-arc graph but also a minimal forbidden induced subgraph for the class of normal Helly circular-arc graphs. In what follows, we use \emph{net} and \emph{tent} as shorthands for $2$-net and $3$-tent, respectively.

If a graph $G$ is the intersection graph of a finite family $\mathcal F$, then $\mathcal F$ is called an \emph{intersection model} of $\mathcal F$. Moreover, if $\mathcal F$ consists of intervals on a line, then $\mathcal F$ is called an \emph{interval model} of $G$, whereas if $\mathcal F$ consists of arcs on a circle, then $\mathcal F$ is called a \emph{circular-arc model} of $G$. Let $G$ be an interval graph containing sets of vertices $A$ and $B$. The pair $A,B$ is said \emph{left-right} if $G$ has an interval model where all intervals corresponding to vertices of $A$ have the same left endpoint and no other endpoints are further to the left and all vertices of $B$ have the same right endpoint and no other endpoints are further to the right \citep{F-G-M-S-sourcesink}. A vertex $v$ of $G$ is an \emph{end} of $G$ if the pair $\{v\},\emptyset$ is left-right. \cite{Gim-endvertex} gave the following characterization of end vertices.

\begin{thm}[\citealp{Gim-endvertex}]\label{thm:Gimbel} Let $G$ be an interval graph. If $v$ is a vertex of $G$, then $v$ is an end vertex of $G$ if and only if $G$ contains none of the graphs in Figure~\ref{fig:end-vertex} as an induced subgraph where the filled vertex represents $v$.
\begin{figure}[!ht]
\centering
\includegraphics{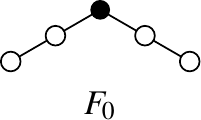}\quad\qquad\includegraphics{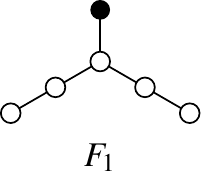}\quad\qquad\includegraphics{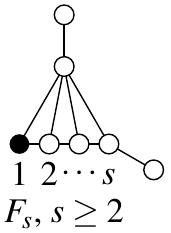}
\caption{Forbidden induced subgraphs characterizing end vertices. Filled vertices are not end vertices}\label{fig:end-vertex}
\end{figure}
\end{thm}
Left-right pairs were characterized by \citet{F-G-M-S-sourcesink}, as follows.
\begin{thm}[\citealp{F-G-M-S-sourcesink}]\label{thm:F-G-M-S-sourcesink} Let $G$ be an interval graph and $A,B\subseteq V(G)$. Then, $A,B$ is left-right  if and only if each of the following assertions holds:
\begin{enumerate}
 \item $A$ and $B$ both are cliques.
 \item Each vertex of $A\cup B$ is an end vertex.
 \item Given any pair of vertices $u$ and $v$, both in $A$ or both in $B$, there is no any chordless path on four vertices in $G$, with $u$ and $v$ as internal vertices of the path;
 \item For each $a \in A$ and $b\in B$, there is no chordless $a,b$-path in $G$, together with a vertex $v$ which is adjacent in $G$ with no vertex on the path.
\end{enumerate}\end{thm}

In the next section, we will make use of the theorem below, whose proof is an easy adaptation of the proof of Theorem~6 in \cite{B-D-G-S-circ-arc}. We give the adapted proof for completeness.

\begin{thm}[adapted from Theorem~6 of \citealp{B-D-G-S-circ-arc}]\label{thm:technical} Let $G$ be a graph containing no induced $G_1$, $G_2$, $G_3$, $G_4$, \text{domino}, $k$-wheel for any $k\geq 4$, or $C_k^*$ for any $k\geq 4$. If $C$ is a chordless cycle of $G$ and $v\in V(G)\setminus V(C)$, then the neighbors of $v$ in $V(C)$ induce a chordless path in $G$.\end{thm}
\begin{myproof} First, notice that $N_G(v)\cap V(C)\neq\emptyset$ and $N_G(v)\cap V(C)\neq V(C)$ because $G$ contains no induced $k$-wheel and no induced $C_k^*$ for any $k\geq 4$. Suppose, by the way of contradiction, that the neighbors of $v$ in $V(C)$ do not induce a chordless path in $G$. Then, the nonneighbors of $v$ in $V(C)$ do not induce a connected subgraph in $G$ and let $P^1$ and $P^2$ be two components of the subgraph of $G$ induced by the nonneighbors of $v$ in $V(C)$. By construction, $P^1$ and $P^2$ are chordless paths and, by symmetry, assume that $\vert V(P^1)\vert\geq\vert V(P^2)\vert$. Let $x_1$ and $x_2$ (resp.\ $y_1$ and $y_2$) be the neighbors of the endpoints of $P^1$ (resp.\  $P^2$) in $V(C)$. Without loss of generality, assume that $x_1,y_1,x_2,y_2$ are labeled in such a way that, in the subgraph of $G$ induced by $V(C)\setminus(V(P^1)\cup V(P^2))$, $x_1$ and $y_2$ are in the same component and also $x_2$ and $y_1$ are in the same component. Let $k=\vert V(P^2)\vert+3$; clearly, $k\geq 4$.

Suppose first that $x_1=y_2$ and $x_2\neq y_1$. On the one hand, if $\vert V(P^1)\vert=1$, then also $\vert V(P^2)\vert=1$ and $V(P^1)\cup V(P^2)\cup\{x_1,x_2,y_1,v\}$ would induce domino or $G_3$, depending on whether $\vert V(C)\vert\geq 6$ or not, respectively, a contradiction. On the other hand, if $\vert V(P^1)\vert\geq 2$, then $V(P^2)\cup\{y_1,y_2,v,x\}$ would induce $C_k^*$ for any vertex $x$ of $P^1$ nonadjacent to $x_1$, a contradiction. These contradictions show, by symmetry, that either $x_1=y_2$ and $x_2=y_1$, or $x_1\neq y_2$ and $x_2\neq y_1$. Notice that we can assume that $\vert V(P^1)\vert\leq 2$ because otherwise $V(P^2)\cup\{y_1,y_2,v,x\}$ would induce $C_k^*$ in $G$ for any vertex $x$ in $P^1$ nonadjacent to both $x_1$ and $x_2$. Therefore, if $x_1=y_2$ and $x_2=y_1$, then $V(C)\cup\{v\}$ would induce $G_1$, or $G_2$, or $G_4$ in $G$, a contradiction. Finally, if $x_1\neq y_2$ and $x_2\neq y_1$, then $V(P^2)\cup\{y_1,y_2,v,x\}$ would induce $C_k^*$ for any vertex $x$ in $P^1$, a contradiction. These contradictions arose from assuming that the neighbors of $v$ in $V(C)$ do not induce a chordless path in $V(C)$.\end{myproof}

\section{Forbidden induced subgraph characterization of normal Helly circular-arc graphs}\label{sec:main}

The main result of this section is Theorem~\ref{thm:main} which characterizes normal Helly circular-arc graphs by minimal forbidden induced subgraphs. \cite*{L-S-S-normal} solved this problem partially, by restricting themselves to circular-arc graphs; i.e., they found the list of all those circular-arc graphs which are minimal forbidden induced subgraphs for the class of normal Helly circular-arc graphs, as follows.

\begin{thm}[\citealp{L-S-S-normal}]\label{thm:Sou-HNCA} Let $H$ be a circular-arc graph. If $H$ is a minimal forbidden induced subgraph for the class of normal Helly circular-arc graphs, then $H$ is isomorphic to one of the following graphs: umbrella, net, $k$-tent for some $k\geq 3$, or $k$-wheel some $k\geq 4$.\end{thm}

The remaining of this section is devoted the state and prove our main result below.

\begin{thm}\label{thm:main} A graph $G$ is a normal Helly circular-arc graph if and only if $G$ contains no induced bipartite claw, umbrella, $k$-net for any $k\geq 2$, $k$-tent for any $k\geq 3$, $k$-wheel for any $k\geq 4$, $G_1$, $G_2$, $G_3$, $G_4$, domino, $\overline{C_6}$, or $C_k^*$ for any $k\geq 4$ (see~Figures~\ref{fig:LekkerkerkerBoland} and \ref{fig:technical}).\end{thm}
\begin{myproof} The necessity is clear. So, assume that $G$ contains no induced bipartite claw, umbrella, $k$-net for any $k\geq 2$, $k$-tent for any $k\geq 3$, $k$-wheel for any $k\geq 4$, $G_1$, $G_2$, $G_3$, $G_4$, domino, $\overline{C_6}$, or $C_k^*$ for any $k\geq 4$. Because of Theorem~\ref{thm:Sou-HNCA}, in order to prove that $G$ is a normal Helly circular-arc graph, it suffices to show that $G$ is a circular-arc graph. If $G$ is chordal, then $G$ is an interval graph by Theorem~\ref{thm:Lekkerkerker-Boland} and, in particular, a circular-arc graph. Thus, we assume, without loss of generality, that $G$ has some chordless cycle $C=v_1v_2\ldots v_nv_1$ for some $n\geq 4$. In what follows, vertex and set subindices should be understood modulo $n$.

For each $i\in\{1,\ldots,n\}$, we define the sets $A_i$, $B_i$, $O_i$, and $T_i$ as follows:
\begin{itemize}
 \item A vertex $v\in V(G)$ belongs to $A_i$ if and only if $v$ is adjacent to $v_{i-1}$ and $v_i$ and also adjacent to some vertex $w$ which is adjacent to $v_i$ and nonadjacent to $v_{i-1}$. (For instance, $w$ may be~$v_{i+1}$.)
 \item A vertex $v\in V(G)$ belongs to $B_i$ if and only if $v$ is adjacent to $v_i$ and $v_{i+1}$ an also adjacent to some vertex $w$ which is adjacent to $v_i$ and nonadjacent to $v_{i+1}$. (For instance, $w$ may be $v_{i-1}$.)
 \item A vertex $v\in V(G)$ belongs to $O_i$ if and only if $N_G(v)\cap V(C)=\{v_i\}$.
 \item A vertex $v\in V(G)$ belongs to $T_i$ if and only if $N_G(v)\cap V(C)=\{v_i,v_{i+1}\}$ and $v\notin B_i\cup A_{i+1}$. 
\end{itemize}

Claims~\ref{claim:vertices} to \ref{claim:neigh-C} below will be used to build a circular-arc model for $G$; see paragraph \emph{`Constructing a circular-arc model for $G$'} immediately after the proof of Claim~\ref{claim:neigh-C}.

\begin{claim}\label{claim:vertices} $V(G)\setminus V(C)=\bigcup_{i=1}^n (A_i\cup B_i\cup O_i\cup T_i)$.\end{claim}
\begin{claimproof} Since $C$ is a chordless cycle, it is clear by definition that no vertex in $\bigcup_{i=1}^n (A_i\cup B_i\cup O_i\cup T_i)$ can belong to $V(C)$. Conversely, let $v\in V(G)\setminus V(C)$. Since $G$ contains no induced $C_n^*$, $v$ has at least one neighbor in $V(C)$. If $\vert N_G(v)\cap V(C)\vert=1$, then $N_G(v)\cap V(C)=\{v_i\}$ for some $i\in\{1,2,\ldots,n\}$ and, by definition, $v\in O_i$. If $\vert N_G(v)\cap V(C)\vert=2$, then, by Theorem~\ref{thm:technical}, $N_G(v)\cap V(C)=\{v_i,v_{i+1}\}$ for some $i\in\{1,2,\ldots,n\}$ and, by definition, $v\in T_i\cup B_i\cup A_{i+1}$. Finally, if $\vert N_G(v)\cap V(C)\vert\geq 3$, then, by Theorem~\ref{thm:technical}, $v$ is adjacent to $v_{i-1}$, $v_i$, and $v_{i+1}$ for some $i\in\{1,2,\ldots,n\}$ and, by definition,~$v\in A_i$.\end{claimproof}

In each of the claims below, $i$ is any integer belonging to $\{1,2,\ldots,n\}$.
\begin{claim}\label{claim:oadjacent} Every vertex $v$ of $G$ which is adjacent simultaneously to $v_{i-1}$, $v_i$, and $v_{i+1}$, is also adjacent to every vertex $o$ in $O_i$.\end{claim}
\begin{claimproof} As $v$ is adjacent to $v_{i-1}$, $v_i$, and $v_{i+1}$, Theorem~\ref{thm:technical} implies that $N_G(v)\cap V(C)$ induces a chordless path $P=v_pv_{p+1}\ldots v_{p+m}$ for some $p\in\{1,\ldots,n\}$ and some $m\leq n-2$ such that $v_i$ is an interior vertex of $P$. Thus, if $v$ were nonadjacent to some $o\in O_i$, then $\{v,v_{p+m},v_{p+m+1},\ldots,v_p,o\}$ would induce $C_{n-m+2}^*$ in $G$, where $n-m+2\geq 4$, a contradiction.  This contradiction proves the~claim.\end{claimproof}

\begin{claim}\label{claim:intervals} $G[A_i\cup O_i\cup B_i]$ and $G[B_i\cup T_i\cup A_{i+1}]$ are interval graphs.\end{claim}
\begin{claimproof} By Theorem~\ref{thm:Lekkerkerker-Boland}, it suffices to prove that $G[A_i\cup O_i\cup B_i]$ and $G[B_i\cup T_i\cup A_{i+1}]$ contain no induced $C_k$ for any $k\geq 4$. If there were any chordless cycle $C'$ in $G[A_i\cup O_i\cup B_i]$ or $G[B_i\cup T_i\cup A_{i+1}]$, then, since every vertex in $A_i\cup O_i\cup B_i\cup T_i\cup A_{i+1}$ is adjacent to $v_i$, $V(C')\cup\{v_i\}$ would induce $k$-wheel in $G$ where $k=\vert V(C')\vert\geq 4$, a contradiction. This contradiction proves the claim.\end{claimproof}

Claims~\ref{claim:aandbcliques} to \ref{claim:aibiendvertices} below, together with Theorem~\ref{thm:F-G-M-S-sourcesink} and Claim~\ref{claim:intervals}, will prove that $A_i,B_i$ is a left-right pair of $G[A_i\cup O_i\cup B_i]$ (Claim~\ref{claim:left-right-AOB}).

\begin{claim}\label{claim:aandbcliques} $A_i$ and $B_i$ are cliques of $G$.\end{claim}
\begin{claimproof} Let $a_1,a_2\in A_i$ such that $a_1\neq a_2$ and assume, by the way of contradiction, that $a_1$ and $a_2$ are nonadjacent. By definition, for each $j\in\{1,2\}$, there is some neighbor $w_j$ of $a_j$ such that $w_j$ is adjacent to $v_i$ and nonadjacent to $v_{i-1}$. Notice that $a_2$ is nonadjacent to $w_1$, since otherwise $\{v_i,v_{i-1},a_1,w_1,a_2\}$ would induce $4$-wheel in $G$. Symmetrically, $a_1$ is nonadjacent to $w_2$ and, necessarily, $w_1\neq w_2$. Moreover, $w_1$ is nonadjacent to $w_2$, since otherwise $\{v_i,v_{i-1},a_1,w_1,w_2,a_2\}$ would induce $5$-wheel in $G$. We notice that $a_1$ and $a_2$ cannot be simultaneously adjacent to $v_{i-2}$, since otherwise $\{v_{i-1},v_i,a_1,v_{i-2},a_2\}$ would induce $4$-wheel in $G$. Without loss of generality, assume that $a_2$ is nonadjacent to $v_{i-2}$. If $a_1$ were adjacent to $v_{i-2}$, then either $\{v_i,v_{i-1},v_{i-2},w_1,a_1\}$ would induce $4$-wheel in $G$ or $\{v_i,v_{i-1},a_1,a_2,v_{i-2},w_1\}$ would induce tent in $G$, depending on whether $w_1$ is adjacent to $v_{i-1}$ or not, respectively. Therefore, also $a_1$ is nonadjacent to $v_{i-2}$. Notice that at least one of $w_1$ and $w_2$ is adjacent to $v_{i-2}$, since otherwise $\{v_{i-2},w_1,a_1,v_{i-1},a_2,w_2,v_i\}$ would induce umbrella in $G$. If both $w_1$ and $w_2$ are adjacent to $v_{i-2}$, then $\{v_i,v_{i-1},v_{i-2},w_1,w_2\}$ induces $G_1$ in $G$, whereas if only $w_1$ is adjacent to $v_{i-2}$, then $\{w_1,v_{i-2},v_{i-1},a_1,w_2\}$ induces $C_4^*$ in $G$. These contradictions arose from assuming that $a_1$ and $a_2$ were nonadjacent. We conclude that $A_i$ is a clique of $G$ and, by symmetry, that also $B_i$ is a clique of $G$.\end{claimproof}

\begin{claim}\label{claim:P4-AOB} There are no two vertices $u$ and $v$, both in $A_i$ or both in $B_i$, such that $u$ and $v$ are the internal vertices of a chordless path on four vertices in $G[A_i\cup O_i\cup B_i]$.\end{claim}
\begin{claimproof} Suppose, by the way of contradiction, that the claim is false; i.e., there are two vertices $u$ and $v$, both in $A_i$ or both in $B_i$, such that $u$ and $v$ are the internal vertices of a chordless path $P$ on four vertices in $G[A_i\cup O_i\cup B_i]$. By symmetry, we assume, without loss of generality, that $u$ and $v$ both belong to $A_i$ and let $a_1=u$ and $a_2=v$. As we proved in Claim~\ref{claim:aandbcliques} that $A_i$ and $B_i$ are cliques, the endpoints of $P$ either both belong to $O_i$ or one belongs to $O_i$ and the other one to $B_i$. Let us assume, without loss of generality, that $P=xa_1a_2o$ where $o\in O_i$ and either $x\in O_i$ or $x\in B_i$.

We assert that $x$ is adjacent to $v_i$ and nonadjacent to both $v_{i-1}$ and $v_{i-2}$. The assertion is clearly true if $x\in O_i$; so, in what remains of this paragraph, we assume, without loss of generality, that $x\in B_i$. By definition, $x$ is adjacent to $v_i$ and $v_{i+1}$ and, by tion, nonadjacent to $o$. Hence, Claim~\ref{claim:oadjacent} implies $x$ is nonadjacent to $v_{i-1}$. Moreover, $x$ is nonadjacent to $v_{i-2}$, since otherwise either $\{a_1,v_{i-2},v_{i-1},v_i,x\}$ would induce $4$-wheel in $G$ or $\{v_{i-2},v_{i-1},a_1,x,o\}$ would induce $C_4^*$ in $G$, depending on whether $a_1$ is adjacent to $v_{i-2}$ or not, respectively. This proves the assertion.

If none of $a_1$ and $a_2$ were adjacent to $v_{i-2}$, then $\{v_{i-2},v_{i-1},x,a_1,a_2,o\}$ would induce net in $G$. If both $a_1$ and $a_2$ were adjacent to $v_{i-2}$, then $\{v_{i-2},v_i,x,a_1,a_2,o\}$ would induce tent in $G$. Finally, if exactly one of $a_1$ and $a_2$ were adjacent to $v_{i-2}$, then $\{v_i,a_1,v_{i-1},a_2,x,o,v_{i-2}\}$ would induce $4$-tent in $G$. These contradictions prove the claim.\end{claimproof}

\begin{claim}\label{claim:abpath-AOB} For each $a\in A_i$ and each $b\in B_i$, there is no chordless $a,b$-path in $G[A_i\cup O_i\cup B_i]$ together with a vertex $v\in A_i\cup O_i\cup B_i$ such that $v$ is adjacent to no vertex of the path.\end{claim}
\begin{claimproof} Suppose, by the way of contradiction, that the claim is false and let $P=x_0x_1\ldots x_p$ be a chordless path in $G[A_i\cup O_i\cup B_i]$ with minimum number of vertices such that $x_0\in A_i$, $x_p\in B_i$, and there is a vertex $v\in A_i\cup O_i\cup B_i$ such that $v$ is adjacent to no vertex of $P$. As $A_i$ and $B_i$ are cliques (by Claim~\ref{claim:aandbcliques}), $V(P)\cap A_i\subseteq\{x_0,x_1\}$ and $V(P)\cap B_i\subseteq\{x_{p-1},x_p\}$. Moreover, if $x_1\in A_i$, then $P'=x_1x_2\ldots x_p$ would be a chordless path in $G[A_i\cup O_i\cup B_i]$ having less vertices than $P$ and such that $x_1\in A_i$, $x_p\in B_i$, and $v$ would be adjacent to no vertex of $P'$, which would contradict the choice of $P$. Hence, $x_0$ is the only vertex of $A_i$ in $P$ and, symmetrically, $x_p$ is the only vertex of $B_i$ in $P$; i.e., $x_1,x_2,\ldots,x_{p-1}\in O_i$. Since $A_i$ and $B_i$ are cliques and since $v$ is nonadjacent to $x_0\in A_i$ and $x_p\in B_i$, $v\in O_i$ holds. Let $a=x_1$, $b=x_p$, and $o=v$. Since $a\in A_i$ and $b\in B_i$ are nonadjacent to $o$, Claim~\ref{claim:oadjacent} implies that $a$ is nonadjacent to $v_{i+1}$ and $b$ is nonadjacent to $v_{i-1}$. In particular, $a\neq b$ and $p\geq 1$. Theorem~\ref{thm:technical} guarantees that there are some $q,r\in\{1,\ldots,n\}$ such that $N_G(a)\cap V(C)=\{v_i,v_{i-1},\ldots,v_q\}$ and $N_G(b)\cap V(C)=\{v_i,v_{i+1},\ldots,v_r\}$. If $a$ and $b$ had no common neighbor in $V(C)$ apart from $v_i$, then $V(P)\cup\{v_r,v_{r+1},\ldots,v_q,o\}$ would induce $C_k^*$ in $G$ for some $k\geq 4$, a contradiction. Hence, $a$ and $b$ have some common neighbor in $V(C)$ different from $v_i$ and necessarily $v_r\in N_G(a)$. Notice that $v_r\neq v_i$ because $b$ is adjacent to $v_{i+1}$ and that $v_r\neq v_{i+1}$ because $a$ is adjacent to $v_r$ and nonadjacent to $v_{i+1}$. Therefore, if $p=1$, then $V(P)\cup\{v_i,v_{i-1},\ldots,v_r\}$ induces a $k$-wheel in $G$ for some $k\geq 4$, a contradiction. On the contrary, if $p\geq 2$, then $V(P)\cup\{v_r,o\}$ induces $C_{p+2}^*$ in $G$ where $p+2\geq 4$, a contradiction. These contradictions prove the claim.\end{claimproof}

\begin{claim}\label{claim:aibiendvertices} Each vertex of $A_i\cup B_i$ is an end vertex in $G[A_i\cup O_i\cup B_i]$.\end{claim}
\begin{claimproof} By symmetry, it suffices to prove that each vertex of $A_i$ is an end vertex of $G[A_i\cup O_i\cup B_i]$. Suppose, by the way of contradiction, that not every vertex of $A_i$ is an end vertex of $G[A_i\cup O_i\cup B_i]$. By Theorem~\ref{thm:Gimbel}, there is some minimum nonnegative integer $s$ such that $G[A_i\cup O_i\cup B_i]$ contains an induced $F_s$ (see Figure~\ref{fig:end-vertex}) where the filled vertex is some vertex $a\in A_i$.

\emph{Assume first that $s=0$}; i.e., there is a chordless path $P=x_1x_2x_3x_4x_5$ in $G[A_i\cup O_i\cup B_i]$ where $x_3=a$. Since Claim~\ref{claim:aandbcliques} ensures that $A_i$ and $B_i$ are cliques, either $x_1$ or $x_5$ belongs to $O_i$. By symmetry, assume that $x_1\in O_i$. As $x_4$ and $x_5$ belong to the same component of $G[A_i\cup O_i\cup B_i]\setminus N_G[x_1]$ than $a\in A_i$, Claim~\ref{claim:abpath-AOB} implies that none of $x_4$ and $x_5$ belongs to $B_i$, Necessarily, $x_5\in O_i$ because $x_5$ is nonadjacent to $a\in A_i$ and $A_i$ is a clique (by Claim~\ref{claim:aandbcliques}). Symmetrically, from $x_5\in O_i$ we deduce that $x_2\notin B_i$. By Claim~\ref{claim:P4-AOB} applied to the chordless paths $x_1x_2ax_4$ and $x_2ax_4x_5$, it follows that none of $x_2$ and $x_4$ belongs to $A_i$. We conclude that $x_1,x_2,x_4,x_5\in O_i$ and, consequently, either $\{v_{i-2},v_i\}\cup V(P)$ induces umbrella or $\{v_{i-2},v_{i-1}\}\cup V(P)$ induces bipartite claw, depending on whether $a$ is adjacent to $v_{i-1}$ or not, respectively. These contradictions prove that $s\neq 0$.

\emph{Assume now that $s=1$}; i.e., there is a chordless path $P=x_1x_2x_3x_4x_5$ in $G[A_i\cup O_i\cup B_i]$ such that the only neighbor of $a$ in $P$ is $x_3$. Since $A_i$ and $B_i$ are cliques, either $x_1$ or $x_5$ belongs to $O_i$. Without loss of generality, assume that $x_1\in O_i$. Since $x_3$, $x_4$, and $x_5$ belong to the same component of $G[A_i\cup O_i\cup B_i]\setminus N_G[x_1]$ than $a\in A_i$, Claim~\ref{claim:abpath-AOB} implies that none of $x_3$, $x_4$, and $x_5$ belongs to $B_i$. Notice that $x_3\notin A_i$ (since otherwise $s$ would be $0$) and none of $x_4$ and $x_5$ belongs to $A_i$ because both vertices are nonadjacent to $a$ and $A_i$ is a clique (by Claim~\ref{claim:aandbcliques}). Necessarily, $x_3,x_4,x_5\in O_i$. Symmetrically, from $x_5\in O_i$, it follows that $x_1,x_2\in O_i$. Thus, $V(P)\subseteq O_i$, which means that $\{v_{i-1},a\}\cup V(P)$ induces a bipartite claw in $G$, a contradiction. This contradiction proves that $s\neq 1$.

\emph{It only remains to consider the case $s\geq 2$}; i.e., there is a chordless path $P=ax_2\ldots x_sx_{s+1}$ in $G[A_i\cup O_i\cup B_i]$ and two vertices $y_1,y_2\in A_i\cup O_i\cup B_i$ whose neighborhoods in $A_i\cup O_i\cup B_i$ are $\{y_2\}$ and $\{a,x_2,\ldots,x_s\}$, respectively. As $A_i$ and $B_i$ are cliques, it holds that $y_1,x_{s+1}\in O_i\cup B_i$ and at least one of $y_1$ and $x_{s+1}$ belongs to $O_i$. Since $y_1,y_2,x_2,x_3,\ldots,x_{s-1}$ belong to the same component of $G[A_i\cup O_i\cup B_i]\setminus N_G[x_{s+1}]$ than $a$ and $x_2,x_3,\ldots,x_{s+1}$ belong to the same component of $G[A_i\cup O_i\cup B_i]\setminus N_G[y_1]$ than $a$, Claim~\ref{claim:abpath-AOB} implies that the fact that at least one of $y_1$ and $x_{s+1}$ belongs to $O_i$ means that none of $x_2$, $x_3$, \ldots, $x_s$, $x_{s+1}$, $y_1$, and $y_2$ belongs to $B_i$. Hence, $x_2,y_2\in A_i\cup O_i$ and, as $A_i$ is a clique, $x_3,x_4,\ldots,x_{s+1},y_1\in O_i$. Notice that at least one of $x_2$ and $y_2$ belongs to $A_i$, since otherwise $\{v_{i-1},a,x_2,x_3,\ldots,x_s,y_1,y_2\}$ would induce $s$-net in $G$.

Assume first that $x_2\in A_i$ but $y_2\notin A_i$. On the one hand, if $s\geq 3$, then $\{v_{i-1},x_2,x_3,$ %line-break
$\ldots,x_{s+1},y_1,y_2\}$ would induce $(s-1)$-net in $G$. On the other hand, if $s=2$, then the following assertions hold: if $a$ and $x_2$ were adjacent to $v_{i-2}$, then $\{v_{i-2},v_i,a,x_2,x_3,y_1,y_2\}$ would induce $4$-tent in $G$; if $a$ and $x_2$ were nonadjacent to $v_{i-2}$, then $\{v_{i-2},v_{i-1},a,x_2,x_3,$  % line-break
$y_1,y_2\}$ would induce $3$-net in $G$; if $a$ were adjacent to $v_{i-2}$ and $x_2$ were nonadjacent to $v_{i-2}$, then $\{v_{i-2},a,x_2,x_3,y_1,y_2\}$ would induce net in $G$; if $a$ were nonadjacent to $v_{i-2}$ but $x_2$ were adjacent to $v_{i-2}$, then $\{v_{i-2},v_{i-1},v_i,a,x_2,x_3,y_1,y_2\}$ would induce $5$-tent in $G$. These contradictions arose from assuming that $x_2\in A_i$ but $y_2\notin A_i$.

Suppose now that $y_2\in A_i$ but $x_2\notin A_i$. If both vertices $a$ and $y_2$ were nonadjacent to $v_{i-2}$, then $\{v_{i-2},v_{i-1},a,x_2,x_3,\ldots,x_{s+1},y_1,y_2\}$ would induce $(s+1)$-net in $G$. If $a$ and $y_2$ were adjacent to $v_{i-2}$, then $\{v_{i-2},v_i,a,x_2,x_3,\ldots,x_{s+1},y_1,y_2\}$ would induce $(s+2)$-tent in $G$.  If $a$ were adjacent to $v_{i-2}$ but $y_2$ were nonadjacent to $v_{i-2}$, then $\{v_{i-2},a,x_2,x_3,\ldots,x_{s+1},y_1,y_2\}$ would induce $s$-net in $G$. If $a$ were nonadjacent to $v_{i-2}$ and $y_2$ were adjacent to $v_{i-2}$, then $\{v_{i-2},v_{i-1},v_i,a,x_2,x_3,\ldots,x_{s+1},y_1,y_2\}$ would induce $(s+3)$-tent in $G$. These contradictions arose from assuming that $y_2\in A_i$ but $x_2\notin A_i$

Necessarily, both $x_2$ and $y_2$ belong to $A_i$. Notice that $s\geq 3$, since otherwise the chordless path $y_1y_2x_2x_3$ in $G[A_i\cup O_i\cup B_i]$ would have both interior vertices in $A_i$, contradicting Claim~\ref{claim:P4-AOB}. If none of $x_2$ and $y_2$ is adjacent to $v_{i-2}$, then $\{v_{i-2},v_{i-1},x_2,x_3,\ldots,$ %line-break
$x_{s+1},y_1,y_2\}$ induces $s$-net in $G$. If both $x_2$ and $y_2$ are adjacent to $v_{i-2}$, then $\{v_{i-2},v_i,x_2,$ %line-break
$x_3,\ldots,x_{s+1},y_1,y_2\}$ induces $(s+1)$-tent in $G$. If, from $x_2$ and $y_2$, only $y_2$ is adjacent to $v_{i-2}$, then $\{v_{i-2},v_{i-1},v_i,x_2,x_3,\ldots,x_{s+1},y_1,y_2\}$ induces $(s+2)$-tent in $G$. Finally, if from $x_2$ and $y_2$, only $x_2$ is adjacent to $v_{i-2}$, then $\{v_{i-2},x_2,x_3,\ldots,x_{s+1},y_1,y_2\}$ induces $(s-1)$-net in $G$. These contradictions arose from assuming that there was some vertex of $A_i$ which was not an end vertex of $G[A_i\cup O_i\cup B_i]$. Thus, every vertex of $A_i$ is an end vertex of $G[A_i\cup O_i\cup B_i]$ and, symmetrically, every vertex of $B_i$ is also an end vertex of $G[A_i\cup O_i\cup B_i]$.\end{claimproof}

\begin{claim}\label{claim:left-right-AOB} $A_i$,$B_i$ is a left-right pair of $G[A_i\cup O_i\cup B_i]$.\end{claim}
\begin{claimproof} It follows from Theorem~\ref{thm:F-G-M-S-sourcesink} and Claims~\ref{claim:intervals} to \ref{claim:aibiendvertices}.\end{claimproof}

\begin{claim}\label{claim:bwt} If some vertex $b\in B_i$ is nonadjacent to some vertex $x\in T_i\cup A_{i+1}$, then $b$ is nonadjacent to $v_{i+2}$ and, for every neighbor $w$ of $b$ such that $w$ is adjacent to $v_i$ and nonadjacent to $v_{i+1}$, $w$ is also nonadjacent to $x$.\end{claim}
\begin{claimproof} Let $b$ be any vertex in $B_i$ being nonadjacent to some vertex $x\in T_i\cup A_{i+1}$. As $b\in B_i$, there is some neighbor $w$ of $b$ such that $w$ is adjacent to $v_i$ and nonadjacent to $v_{i+1}$. Necessarily, $w$ is nonadjacent to $x$, since otherwise $\{v_i,v_{i+1},b,w,x\}$ would induce $4$-wheel in $G$. Suppose, by the way of contradiction, that $b$ were adjacent to $v_{i+2}$. Therefore, by definition, $b\in A_{i+1}$. As $A_{i+1}$ is a clique (by Claim~\ref{claim:aandbcliques}) but $x$ is nonadjacent to $b$, it holds that $x\notin A_{i+1}$. Consequently, $x$ is nonadjacent to $v_{i+2}$. Besides, $w$ is nonadjacent to $v_{i+2}$, since otherwise $\{b,v_i,v_{i+1},v_{i+2},w\}$ would induce a $4$-wheel in $G$. We conclude that, if $b$ were adjacent to $v_{i+2}$, then $\{w,v_i,v_{i+1},v_{i+2},b,x\}$ would induce tent in $G$, a contradiction. This contradiction proves that $b$ is nonadjacent to $v_{i+2}$. As we already proved that $w$ is nonadjacent to $x$, the proof of the claim is complete.\end{claimproof}

Claims \ref{claim:P4-BTA} to \ref{claim:end-vertex-BTA} below, together with Theorem~\ref{thm:F-G-M-S-sourcesink} and Claims~\ref{claim:intervals} and \ref{claim:aandbcliques}, will prove that $B_i,A_{i+1}$ are left-right sets in $G[B_i\cup T_i\cup A_{i+1}]$ (Claim~\ref{claim:left-right-BTA}).

\begin{claim}\label{claim:P4-BTA} There are no two vertices $u$ and $v$, both in $B_i$ or both in $A_{i+1}$, such that $u$ and $v$ are the internal vertices of a chordless path on four vertices in $G[B_i\cup T_i\cup A_{i+1}]$.\end{claim}
\begin{claimproof} Suppose, by the way of contradiction, that there are two vertices $u$ and $v$, both in $B_i$ or both in $A_{i+1}$, such that $u$ and $v$ are the internal vertices of a chordless path $P$ on four vertices in $G[B_i\cup O_i\cup A_{i+1}]$. By symmetry, we assume, without loss of generality, that $u$ and $v$ both belong to $B_i$. As Claim~\ref{claim:aandbcliques} ensures that that $B_i$ and $A_{i+1}$ are cliques, the endpoints of $P$ either both belong to $T_i$ or one belongs to $T_i$ and the other one to $A_{i+1}$. Assume, without loss of generality, that $P=xb_1b_2t$ where $b_1,b_2\in B_i$, $t\in T_i$, and either $x\in A_i$ or $x\in T_i$. For each $j\in\{1,2\}$, let $w_j$ be a neighbor of $b_j$ such that $w_j$ is adjacent to $v_i$ and nonadjacent to $v_{i+1}$. As $b_1\in B_i$ is nonadjacent to $t\in T_i$ and $b_2\in B_i$ is nonadjacent to $x\in T_i\cup A_{i+1}$, Claim~\ref{claim:bwt} implies that $b_1$ and $b_2$ are nonadjacent to $v_{i+2}$, $w_1$ is nonadjacent to $t$, and $w_2$ is nonadjacent to $x$. If $w_1$ were adjacent to $x$, then, by definition of $B_i$, $x\in B_i$, contradicting the facts that $x$ is nonadjacent to $b_2\in B_i$ and $B_i$ is a clique. If $w_2$ were adjacent to $t$, then, by definition of $B_i$, $t\in B_i$, contradicting $t\in T_i$. Therefore, $w_1$ is nonadjacent to $x$ and $w_2$ is nonadjacent to $t$. Notice that $w_1$ is nonadjacent to $b_2$, since otherwise $\{w_1,b_1,b_2,x,v_{i+1},t\}$ would induce tent in $G$. Also $w_2$ is nonadjacent to $b_1$, since otherwise $\{w_2,b_1,b_2,x,v_{i+1},t\}$ would induce tent in $G$. Consequently, $w_1\neq w_2$. Moreover, $w_1$ and $w_2$ are nonadjacent, since otherwise $\{v_i,b_1,b_2,w_2,w_1\}$ would induce $4$-wheel in $G$. Recall that, by Claim~\ref{claim:bwt}, $b_1$ and $b_2$ are nonadjacent to $v_{i+2}$. Notice that at least one of $w_1$ and $w_2$ is adjacent to $v_{i+2}$, since otherwise $\{b_1,b_2,v_{i+1},w_1,w_2,v_{i+2}\}$ would induce net in $G$. If both $w_1$ and $w_2$ were adjacent to $v_{i+2}$, then $\{v_i,v_{i+1},v_{i+2},w_1,w_2\}$ would induce $G_1$ in $G$. If $w_1$ were adjacent to $v_{i+2}$ but $w_2$ were nonadjacent to $v_{i+2}$, then $\{b_1,w_1,v_{i+2},v_{i+1},w_2\}$ would induce $C_4^*$ in $G$. If $w_2$ were adjacent to $v_{i+2}$ but $w_1$ were nonadjacent to $v_{i+2}$, then $\{b_2,w_2,v_{i+2},v_{i+1},w_1\}$ would induce $C_4^*$. As these contradictions arose from assuming the existence of $P$, the proof of the claim is complete.\end{claimproof}

\begin{claim}\label{claim:abpath-BTA} For each $b\in B_i$ and each $a\in A_{i+1}$, there is no chordless $a,b$-path in $G[B_i\cup T_i\cup A_{i+1}]$ together with a vertex $v\in A_{i+1}\cup T_i\cup B_i$ such that $v$ is adjacent to no vertex of the path.\end{claim}
\begin{claimproof} Suppose, by the way of contradiction, that the claim is false and let $P=x_0x_1\ldots x_p$ be a chordless path in $G[B_i\cup O_i\cup A_{i+1}]$ with minimum number of vertices such that $x_0\in B_i$, $x_p\in A_{i+1}$, and there is a vertex $v\in B_i\cup O_i\cup A_{i+1}$ which is adjacent to no vertex of $P$. Reasoning in a way entirely analogous to that employed in the beginning of the proof of Claim~\ref{claim:abpath-AOB}, it follows that $x_0$ is the only vertex of $P$ in $B_i$, $x_p$ is the only vertex of $P$ in $A_{i+1}$, and $v,x_1,x_2,\ldots,x_{p-1}\in T_i$. Let $b=x_0$, $a=x_p$, and $t=v$. As $b\in B_i$ and $a\in A_{i+1}$, vertex $b$ has some neighbor $w_1$ adjacent to $v_i$ and nonadjacent to $v_{i-1}$ and $a$ has some neighbor adjacent to $v_{i+1}$ and nonadjacent to $v_i$. By Claim~\ref{claim:bwt}, $w_1$ is nonadjacent to $t$ and, symmetrically, $w_2$ is nonadjacent to $t$. Moreover, $b$ is nonadjacent to $w_2$, since otherwise either $\{b,v_i,v_{i+1},w_2,w_1\}$ would induce $4$-wheel in $G$ or $\{w_1,b,w_2,v_i,v_{i+1},t\}$ would induce tent in $G$, depending on whether $w_1$ is adjacent to $w_2$ or not, respectively. Symmetrically, $a$ is nonadjacent to $w_1$. In particular, $a\neq b$ and $p\geq 1$. Notice that, as $x_1,x_2,\ldots,x_{p-1}\in T_i$, both $w_1$ and $w_2$ are nonadjacent to each interior vertex $P$. We conclude that either $V(P)\cup\{t,w_1,w_2\}$ induces $C_{p+3}^*$ in $G$ or $V(P)\cup\{t,v_i,v_{i+1},w_1,w_2\}$ induces $(p+3)$-tent in $G$, depending on whether $w_1$ and $w_2$ are adjacent or not, respectively. These contradictions show that the claim must be true.\end{claimproof}

\begin{claim}\label{claim:end-vertex-BTA} Each vertex of $B_i\cup A_{i+1}$ is an end vertex in $G[B_i\cup T_i\cup A_{i+1}]$.\end{claim}
\begin{claimproof} By symmetry, it suffices to prove that each vertex of $B_i$ is an end vertex of $G[B_i\cup T_i\cup A_{i+1}]$. Suppose, by the way of contradiction, that not every vertex of $B_i$ is an end vertex of $G[B_i\cup T_i\cup A_{i+1}]$. By Theorem~\ref{thm:Gimbel}, there is some minimum nonnegative integer $s$ such that $G[B_i\cup T_i\cup A_{i+1}]$ contains an induced $F_s$ (see~Figure~\ref{fig:end-vertex}) where the filled vertex is some $b\in B_i$. Since $b\in B_i$, there is some neighbor $w$ of $b$ such that $w$ is adjacent to $v_i$ and nonadjacent to $v_{i+1}$.

\emph{Assume first that $s=0$}; i.e., there is a chordless path $P=x_1x_2x_3x_4x_5$ in $G[B_i\cup T_i\cup A_{i+1}]$ where $x_3=b$. Reasoning in a way entirely analogous to that of case $s=0$ of Claim~\ref{claim:aibiendvertices} (replacing Claims~\ref{claim:P4-AOB} and \ref{claim:abpath-AOB} with Claims~\ref{claim:P4-BTA} and \ref{claim:abpath-BTA}), $x_1,x_2,x_4,x_5\in T_i$ and, by definition of $T_i$, none of $x_1$, $x_2$, $x_4$, and $x_5$ is adjacent to $w$. Hence, $\{x_1,x_2,b,x_4,x_5,v_{i+1},$ $w\}$ induces umbrella in $G$, a contradiction. Thus, $s\neq 0$.

\emph{Assume now that $s=1$}; i.e., there is a chordless path $P=x_1x_2x_3x_4x_5$ in $G[B_i\cup T_i\cup A_{i+1}]$ such that the only neighbor of $b$ in $P$ is $x_3$. Since $B_i$ and $A_{i+1}$ are cliques, either $x_1$ or $x_5$ belongs to $T_i$. Without loss of generality, assume that $x_1\in T_i$. Since $x_3$, $x_4$, and $x_5$ belong to the same component of $G[B_i\cup T_i\cup A_{i+1}]\setminus N_G[x_1]$ than $b\in B_i$, Claim~\ref{claim:abpath-BTA} implies that none of $x_3$, $x_4$, and $x_5$ belongs to $A_{i+1}$. Notice that $x_3\notin B_i$ (since otherwise $s$ would be $0$) and none of $x_4$ and $x_5$ belongs to $B_i$ because both vertices are nonadjacent to $b$ and $B_i$ is a clique (Claim~\ref{claim:aandbcliques}). Necessarily, $x_3,x_4,x_5\in T_i$. Symmetrically, $x_5\in T_i$ implies that $x_1,x_2\in T_i$. Hence, $V(P)\subseteq T_i$ and, by definition of $T_i$, no vertex of $P$ is adjacent to $w$. Consequently, $V(P)\cup\{b,w\}$ induces bipartite claw in $G$, a contradiction. This contradiction proves that $s\neq 1$.

\emph{Since $s\geq 2$}, there are a chordless path $P=ax_2\ldots x_sx_{s+1}$ in $G[B_i\cup T_i\cup A_{i+1}]$ and two vertices $y_1,y_2\in B_i\cup T_i\cup A_{i+1}$ whose neighborhoods in $B_i\cup T_i\cup A_{i+1}$ are $\{y_2\}$ and $\{a,x_2,\ldots,x_s\}$, respectively. As $B_i$ and $A_{i+1}$ are cliques, at least one of $y_1$ and $x_{s+1}$ belongs to $T_i$. Reasoning in a way entirely analogous to that of case $s\geq 2$ of Claim~\ref{claim:aibiendvertices} (replacing Claim~\ref{claim:abpath-AOB} with Claim~\ref{claim:abpath-BTA}), we can prove that $x_3,x_4,\ldots,x_{s+1},y_1\in T_i$ and $x_2,y_2\in B_i\cup T_i$. Notice that at least one of $x_2$ and $y_2$ is adjacent to $w$, since otherwise $\{w,b,x_2,x_3,\ldots,x_{s+1},y_1,y_2\}$ would induce $s$-net in $G$. If $x_2$ is adjacent to $w$ but $y_2$ is nonadjacent to $w$, then either $\{w,x_2,x_3,\ldots,x_{s+1},y_1,y_2\}$ induces $(s-1)$-net in $G$ or $\{w,b,x_2,x_3,y_1,y_2,v_{i+1}\}$ induces $4$-tent in $G$, depending on whether $s\geq 3$ or not, respectively. If $y_2$ is adjacent to $w$ but $x_2$ is nonadjacent to $w$, then $\{w,b,x_2,x_3,\ldots,x_{s+1},y_1,y_2,v_{i+1}\}$ induces $(s+2)$-tent in $G$. Finally, if both $x_2$ and $y_2$ are adjacent to $w$, then $\{w,x_2,x_3,\ldots,x_{s+1},y_1,y_2,v_{i+1}\}$ induces $(s+1)$-tent in $G$. These contradictions arose from assuming that there was some vertex in $B_i$ that was not an end vertex of $G[B_i\cup T_i\cup A_{i+1}]$. This completes the proof of the claim.\end{claimproof}

\begin{claim}\label{claim:left-right-BTA} $B_i,A_{i+1}$ is a left-right pair of $G[B_i\cup T_i\cup A_{i+1}]$.\end{claim}
\begin{claimproof} If follows from Theorem~\ref{thm:F-G-M-S-sourcesink} and Claims~\ref{claim:intervals}, \ref{claim:aandbcliques}, and \ref{claim:P4-BTA} to \ref{claim:end-vertex-BTA}.\end{claimproof}

\begin{claim}\label{claim:consecutivos} If $u$ and $v$ are two adjacent vertices in $V(G)\setminus V(C)$, then $u$ and $v$ have at least one common neighbor in $V(C)$.\end{claim}
\begin{claimproof} Suppose, by the way of contradiction, that there are two adjacent vertices $u,v\in V(G)\setminus V(C)$ having no common neighbor in $V(C)$. By Theorem~\ref{thm:technical}, $N_G(u)\cap V(C)=\{v_i,v_{i+1},\ldots,v_j\}$ and $N_G(v)\cap V(C)=\{v_k,v_{k+1},\ldots,v_\ell\}$, for some $i,j,k,\ell\in\{1,\ldots,n\}$. If $\vert N_G(u)\cap V(C)\vert\geq 3$, then $C'=v_iuv_jv_{j+1}\ldots v_{i-1}v_i$ would be a chordless cycle and the neighbors of $v$ in $V(C')$ would contradict Theorem~\ref{thm:technical}. Hence, $v_j=v_i$ or $v_j=v_{i+1}$. Symmetrically, $v_\ell=v_k$ or $v_\ell=v_{k+1}$. Let $P^1=v_jv_{j+1}\ldots v_k$, $P^2=v_\ell v_{\ell+1}\ldots v_i$. Without loss of generality, $\vert V(P^1)\vert\leq\vert V(P^2)\vert$. Let $m=\vert V(P^1)\vert+2$; hence, $m\geq 4$.

Suppose first that $v_j=v_i$ and $v_\ell=v_k$. Then, $\vert V(P^2)\vert\leq 4$, since otherwise $V(P^1)\cup\{u,v,v_{k+2}\}$ would induce $C_m^*$ in $G$. If $\vert V(P^1)\vert =\vert V(P^2)\vert=4$, then $(V(C)\setminus\{v_k\})\cup\{u,v\}$ would induce bipartite claw in $G$. Therefore, $V(C)\cup\{u,v\}$ induces domino, $G_2$, or $G_4$, depending on whether $(\vert V(P^1)\vert,\vert V(P^2)\vert)=(2,4)$, $(3,3)$, or $(3,4)$, respectively, a contradiction. Thus, $v_j\neq v_i$ or $v_\ell\neq v_k$.

Suppose now that $v_j=v_{i+1}$ and $v_\ell=v_k$. Then, $\vert V(P^2)\vert\leq 3$, since otherwise $V(P^1)\cup\{u,v,v_{k+2}\}$ would induce $C_m^*$ in $G$. If $\vert V(P^1)\vert=\vert V(P^2)\vert=3$, then $(V(C)\setminus\{v_k\})\cup\{u,v\}$ would induce net in $G$. Necessarily, $(\vert V(P^1)\vert,\vert V(P^2)\vert)=(1,2)$ and $V(C)\cup\{u,v\}$ induces $G_3$ in $G$, a contradiction.

Up to symmetry, it only remains to consider the case $v_j=v_{i+1}$ and $v_\ell=v_{k+1}$. On the one hand, if $\vert V(P^2)\vert\geq 3$, then $V(P_1)\cup\{u,v,v_{k+2}\}$ induces $C_m^*$ in $G$. On the other hand, if $\vert V(P^1)\vert=\vert V(P^2)\vert=2$, then $V(C)\cup\{u,v\}$ induces $\overline{C_6}$ in $G$. These contradictions complete the proof of the claim.\end{claimproof}

\begin{claim}\label{claim:neigh-O} If $o\in O_i$, then $N_G(o)=\{v_i\}\cup N_{G[A_i\cup O_i\cup B_i]}(o)$.\end{claim}
\begin{claimproof} Let $o\in O_i$. That $\{v_i\}\cup N_{G[A_i\cup O_i\cup B_i]}(o)\subseteq N_G(o)$ follows easily by definition of $O_i$ and by definition of induced subgraphs. In order to prove the reverse inclusion, let $v\in N_G(o)$. If $v\in V(C)$, then $v=v_i$ by definition of $O_i$; so, assume that $v\notin V(C)$. Since $v\notin V(C)$, we deduce from $ov\in E(G)$ and Claim~\ref{claim:consecutivos} that $v_i\in N_G(v)$. Hence, Theorem~\ref{thm:technical} ensures that $N_G(v)\cap V(C)=\{v_i\}$, $\{v_{i-1},v_i\}\subseteq N_G(v$), or $\{v_i,v_{i+1}\}\subseteq N_G(v)$ and, consequently, $v\in O_i$, $v\in A_i$, or $v\in B_i$, respectively (where $o$ plays the role of $w$ in proving $v\in A_i$ or $v\in B_i$). Thus, $N_G(o)\subseteq\{v_i\}\cup A_i\cup O_i\cup B_i$. Since $o\in O_i$, this is equivalent to $N_G(o)\subseteq\{v_i\}\cup N_{G[A_i\cup O_i\cup B_i]}(o)$, concluding the proof of the claim.
\end{claimproof}

\begin{claim}\label{claim:neigh-T}  If $t\in T_i$, then $N_G(t)=\{v_i,v_{i+1}\}\cup N_{G[B_i\cup T_i\cup A_{i+1}]}(t)$.\end{claim}
\begin{claimproof} Let $t\in T_i$. That $\{v_i,v_{i+1}\}\cup N_{G[B_i\cup T_i\cup A_{i+1}]}\subseteq N_G(t)$ follows immediately by definition of $T_i$ and of induced subgraphs; so, we only need to prove the reverse inclusion. Let $v\in N_G(t)$. If $v\in V(C)$, then $v=v_i$ or $v=v_{i+1}$ by definition of $T_i$; so, assume that $v\notin V(C)$. From Claim~\ref{claim:consecutivos}, Theorem~\ref{thm:technical}, and  the definition of $T_i$, it follows that $N_G(v)\cap V(C)$ contains at least one of the following subsets $\{v_i,v_{i+1}\}$, $\{v_{i-1},v_i\}$, or $\{v_{i+1},v_{i+2}\}$. If $N_G(v)\cap V(C)$ does not contain $\{v_i,v_{i+1}\}$ but one of $\{v_{i-1},v_i\}$ and $\{v_{i+1},v_{i+2}\}$, then $v\in B_i$ or $v\in A_{i+1}$, respectively (where the role of $w$ in proving $v\in B_i$ or $v\in A_{i+1}$ is played by $t$), contradicting $t\in T_i$. This contradiction shows that necessarily $\{v_i,v_{i-1}\}\subseteq N_G(v)$ and, consequently, $v\in A_i\cup T_i\cup B_i$ (as shown in the proof of Claim~\ref{claim:vertices}); thus, since $t\in T_i$, $N_G(t)\subseteq\{v_i,v_{i+1}\}\cup N_{G[B_i\cup T_i\cup A_{i+1}]}$, which completes the proof of the claim.\end{claimproof}

\begin{claim}\label{claim:neigh-C} $N_G(v_i)\setminus V(C)=B_{i-1}\cup T_{i-1}\cup A_i\cup O_i\cup B_i\cup T_i\cup A_{i+1}$.\end{claim}
\begin{claimproof} That $B_{i-1}\cup T_{i-1}\cup A_i\cup O_i\cup B_i\cup T_i\cup A_{i+1}\subseteq N_G(v_i)\setminus V(C)$ follows by definition of the respective sets. In order to prove the reverse inclusion, let $v\in N_G(v_i)\setminus V(C)$. Since $v_i\in N_G(v)$, Theorem~\ref{thm:technical} implies that either $N_G(v)\cap V(C)=\{v_i\}$ or $N_G(v)\cap V(C)$ contains $\{v_{i-1},v_i\}$ or $\{v_i,v_{i+1}\}$ as subsets. On the one hand, if $N_G(v)\cap V(C)=\{v_i\}$, then $v_i\in O_i$ by definition. On the other hand, if $N_G(v)\cap V(C)$ contains $\{v_{i-1},v_i\}$ or $\{v_i,v_{i+1}\}$ as subsets, then $v\in B_{i-1}\cup T_{i-1}\cup A_i$ or $v\in B_i\cup T_i\cup A_{i+1}$, respectively (as shown in the proof of Claim~\ref{claim:vertices}). We conclude that also the reverse inclusion $N_G(v_i)\setminus V(C)\subseteq B_{i-1}\cup T_{i-1}\cup A_i\cup O_i\cup B_i\cup T_i\cup A_{i+1}$ holds and the claim is proved.\end{claimproof}

\smallskip\textbf{Constructing a circular-arc model for $\boldmath G$.}\label{claim:circ-arc} Consider a circular-arc model $\mathcal M$ of the chordless cycle $C$ where 
$\mathcal M$ consists of a set of closed arcs on a circle $\mathcal C$ such that no two arcs of $\mathcal M$ share a common endpoint. For each arc $X$ on $\mathcal C$, we call \emph{left} and \emph{right} endpoints to the starting and ending endpoints of $X$, respectively, when traversing $\mathcal C$ in clockwise direction. For each $j\in\{1,\ldots,n\}$, let $X(v_j)$ be the arc of $\mathcal M$ corresponding to $v_j$ and let $\ell_j$ and $r_j$ be the left and right endpoints of $X(v_j)$, respectively. For each $i\in\{1.\ldots,n\}$, we use Claim~\ref{claim:left-right-AOB} to build an intersection model $\mathcal M^1_i$ of $G[A_i\cup O_i\cup B_{i+1}]$ consisting of open arcs within the clockwise open arc $(r_{i-1},\ell_{i+1})$ on $\mathcal C$ such that the left endpoint of each arc corresponding to a vertex in $A_i$ is $r_{i-1}$ and the right endpoint of each arc corresponding to a vertex in $B_i$ is $\ell_{i+1}$. For each $v\in A_i\cup O_i\cup B_i$, let $X^1_i(v)$ be the arc representing $v$ in $\mathcal M^1_i$. Analogously, for each $i\in\{1,\ldots,n\}$, we use Claim~\ref{claim:left-right-BTA} to build an intersection model $\mathcal M^2_i$ of $G[B_i\cup T_i\cup A_{i+1}]$ consisting of closed arcs within the clockwise closed arc $[\ell_{i+1},r_i]$ on $\mathcal C$ such that the left endpoint of each arc corresponding to a vertex in $B_i$ is $\ell_{i+1}$ and the right endpoint of each arc corresponding to a vertex in $A_{i+1}$ is $r_i$. For each $v\in B_i\cup T_i\cup A_{i+1}$, let $X^2_i(v)$ be the arc representing $v$ in $\mathcal M^2_i$. For each $v\in V(G)\setminus V(C)$, we define $X(v)$\label{def:X(v)} as the union of all the arcs corresponding to $v$ among the models $\mathcal M^1_1,\ldots,\mathcal M^1_n,\mathcal M^2_1,\ldots,\mathcal M^2_n$; i.e., $X(v)=\bigcup_{i:v\in A_i\cup O_i\cup B_i}X^1_i\cup\bigcup_{i:v\in B_i\cup T_i\cup A_{i+1}}X^2_i(v)$.

\vspace{2pt}Claims~\ref{claim:arcs} and \ref{claim:model} prove that the sets $X(v)$ (as defined in the preceding paragraph) are arcs on $\mathcal C$ and that the family consisting of all the arcs $X(v)$ is a circular-arc model for $G$, respectively.
\begin{claim}\label{claim:arcs} For each $v\in V(G)$, $X(v)$ is an arc on $\mathcal C$.\end{claim}
\newcommand{\subL}{_{\,\mathrm L}}
\newcommand{\subR}{_{\,\mathrm R}}
\begin{claimproof} If $v\in V(C)$, the claim is true by definition. Suppose first that $v\in O_i$. Then, since for every $j\in\{1,\ldots,n\}$ each vertex of $A_j\cup O_j\cup B_j$ is adjacent to $v_j$ and each vertex of $B_j\cup T_j\cup A_{j+1}$ is adjacent to $v_j$ and $v_{j+1}$, then $v\in A_j\cup O_j\cup B_j$ only when $j=i$, and $v\notin B_j\cup T_j\cup A_{j+1}$ for each $j\in\{1,\ldots,n\}$; therefore, by definition, $X(v)=X^1_i(v)$. Suppose now that $v\in T_i$. Then, since $v\in B_j\cup T_j\cup A_{j+1}$ only when $j=i$ and $v\notin B_j\cup O_j\cup A_j$ for each $j\in\{1,\ldots,n\}$, the equality $X(v)=X^2_i(v)$ holds. So, assume that $v\in\bigcup_{i=1}^n(A_i\cup B_i)$ and it only remains to show that $X(v)$ is an arc. By Theorem~\ref{thm:technical}, the neighbors of $v$ in $V(C)$ induce a chordless path $P=v_jv_{j+1}\ldots v_k$ for some $j,k\in\{1,\ldots,n\}$ and $j\neq k$. By definition, $v\notin A_i\cup O_i\cup B_i$ for each $i\notin\{j,j+1,\ldots,k\}$ and $v\notin B_i\cup T_i\cup A_{i+1}$ for each $i\notin\{j,j+1,\ldots,k-1\}$. As $v\in A_{j+1},B_{j+1},\ldots,A_{k-1},B_{k-1}$ by definition, the sets $X^1_{j+1}(v)$, $X^2_{j+1}(v)$, \ldots, $X^2_{k-2}(v)$, $X^1_{k-1}(v)$ are the clockwise arcs $(r_j,\ell_{j+2})$, $[\ell_{j+2},r_{j+1}]$, \ldots, $[\ell_{k-1},r_{k-2}]$, $(r_{k-2},\ell_k)$ on $\mathcal C$, respectively, whose union is the clockwise open arc $(r_j,\ell_k)$ on $\mathcal C$. If there is a neighbor $w\subL$ of $v$ adjacent to $v_j$ and nonadjacent to $v_{j+1}$, then $v\in B_j$ and let $X\subL(v)=X^1_j(v)$; otherwise, $v\notin A_j\cup O_j\cup B_j$ and let $X\subL(v)=\emptyset$. If there is a neighbor $w\subR$ of $v$ adjacent to $v_k$ and nonadjacent to $v_{k-1}$, then $v\in A_k$ and let $X\subR(v)=X^1_k(v)$;
otherwise, $v\notin A_k\cup O_k\cup B_k$ and let $X\subR(v)=\emptyset$. Finally, $X(v)=X\subL(v)\cup X^2_j(v)\cup(r_j,\ell_k)\cup X^2_{k-1}(v)\cup X\subR(v)$, which is an arc on $\mathcal C$ because, whenever $X\subL(v)\neq\emptyset$, the right endpoint of $X\subL(v)$ and the left endpoint of $X^2_j(v)$ are $\ell_{j+1}$ and, whenever $X\subR(v)\neq\emptyset$, the right endpoint of $X^2_{k-1}(v)$ and the left endpoint of $X\subR(v)$ are $r_{k-1}$. The proof of the claim is complete.\end{claimproof}

\begin{claim}\label{claim:model} The family consisting of all the arcs $X(v)$ for each $v\in V(G)$ is circular-arc model for~$G$.\end{claim}
\begin{claimproof} Let $\tilde G$ be the graph having the same set of vertices than $G$ and such that two vertices $u$ and $v$ of $\tilde G$ are adjacent if and only if $X(u)$ and $X(v)$ have nonempty intersection. The claim is proved as soon as we verify that $\tilde G=G$. Let $o\in O_j$ for some $j\in\{1,\ldots,n\}$. In the proof of Claim~\ref{claim:arcs} we showed that $X(o)=X^1_j(o)$ which, by construction, intersects precisely the arc $X(v_j)$ and the arcs $X(v)$ for each each neighbor $v$ of $o$ in $A_j\cup O_j\cup B_j$. As a result and by Claim~\ref{claim:neigh-O}, $N_{\tilde G}(o)=N_G(o)$. Let $t\in T_j$ for some $j\in\{1,\ldots,n\}$. In the proof of Claim~\ref{claim:arcs} we showed that $X(t)=X^2_j(t)$ which, by construction, intersects precisely the arcs $X(v_j)$ and $X(v_{j+1})$ and the arcs $X(v)$ for each neighbor of $t$ in $B_j\cup T_j\cup A_{j+1}$. Consequently, by Claim~\ref{claim:neigh-T}, $N_{\tilde G}(t)=N_G(t)$. Notice that, also by construction, for each $j\in\{1,\ldots,n\}$, the arc $X(v_j)$ intersects precisely the arcs $X(v_{j-1})$, $X(v_{j+1})$, and the arcs $X(v)$ for every $v\in B_{j-1}\cup T_{j-1}\cup A_i\cup O_i\cup B_i\cup T_i\cup A_{j+1}$, which, by Claim~\ref{claim:neigh-C}, means that $N_{\tilde G}(v_j)=N_G(v_j)$. In order to complete the proof of the claim, let $u,v\in\bigcup_{i=1}^n(A_i\cup B_i)$ and it only remains to prove that $uv\in E(\tilde G)$ if and only if $uv\in E(G)$. Since $u,v\notin V(C)$, the construction ensures that if $uv\in E(\tilde G)$ then $uv\in E(G)$. For the converse, assume that $uv\in E(G)$. By Claim~\ref{claim:consecutivos} and Theorem~\ref{thm:technical}, $u$ and $v$ have at least one common neighbor in $V(C)$ and the common neighbors of $u$ and $v$ in $V(C)$ induce a chordless path $P$. If $\vert V(P)\vert\geq 2$, then $\{v_j,v_{j+1}\}\subseteq V(P)$ for some $j\in\{1,\ldots,n\}$ and, consequently, $u,v\in B_j\cup T_j\cup A_{j+1}$ (as shown in the proof of Claim~\ref{claim:vertices}); thus, $uv\in E(G[B_j\cup T_j\cup A_{j+1}])$ and, by construction, $uv\in E(\tilde G)$. Finally, let us consider the case $V(P)=\{v_j\}$ for some $j\in\{1,\ldots,n\}$. As $u\notin O_j$ and $v\notin O_j$, then, up to symmetry, $N_G(u)\cap\{v_{j-1},v_j,v_{j+1}\}=\{v_{j-1},v_j\}$ and $N_G(v)\cap\{v_{j-1},v_j,v_{j+1}\}=\{v_j,v_{j+1}\}$. Hence, $u\in A_j$ (with $v$ playing the role of $w$ in the definition) and $v\in B_j$ (with $u$ playing the role of $w$ in the definition); thus, $uv\in E(G[A_j\cup O_j\cup B_j])$ and construction implies that $uv\in E(\tilde G)$. This completes the proof of the claim.\end{claimproof}

By Claim~\ref{claim:model}, $G$ is a circular-arc graph and, as a result, Theorem~\ref{thm:Sou-HNCA} implies that $G$ is a normal Helly circular-arc graph. This completes the proof of Theorem~\ref{thm:main}.\end{myproof}

\bibliographystyle{abbrvnat}
\bibliography{claio2012}

\end{document}